\newif\ifsubmode
\submodefalse
 
\documentclass{emulateapj}
\newcommand{\cg}{$\nabla_{g-r}$}
\newcommand{\cz}{$\nabla_Z$}

\shorttitle{Color gradients}
\shortauthors{La Barbera et al.}

\begin{document}
\title{Color Gradients in Early-type Galaxies: Dependence on
Environment and Redshift}

\author{F. La Barbera\altaffilmark{1}, R.R. de Carvalho\altaffilmark{2}
, R.R. Gal\altaffilmark{3},
G. Busarello\altaffilmark{1},
P. Merluzzi\altaffilmark{1},
M. Capaccioli\altaffilmark{1},
S.G. Djorgovski\altaffilmark{4}
}
\altaffiltext{1}{ INAF -- Osservatorio Astronomico di Capodimonte,
Napoli, Italy, labarber@na.astro.it}
\altaffiltext{2}{ INPE/MCT, Avenida dos Astronautas, 1758,
S$\tilde{a}$o Jos\' e dos
Campos, SP 12227-010, Brazil } 
\altaffiltext{3}{ Department of Physics, University of California -- Davis, One Shields Avenue, Davis, CA 95616}
\altaffiltext{4}{ Department of Physics, Mathematics, and Astronomy, California Institute of Technology, MS 105-24, Pasadena, CA 91125} 

\begin{abstract}
Color gradients  in early-type  galaxies contain valuable  clues about
their formation and evolutionary histories and mechanisms.  We examine
color gradients  in 1,700 early-type  galaxies in 159  galaxy clusters
spanning  a  redshift  range of  0.05  to  0.2.   We find  that  color
gradients strongly  depend on  the environment where  galaxies reside,
with steeper  color gradients in  poor rather than rich  clusters.  No
dependence of color gradients on  galaxy luminosity is found in either
rich  or poor  clusters.  The  difference  in color  gradients can  be
explained  by a  change  in  the internal  metallicity  and/or an  age
gradient in  these galaxies. Our results support  a reasonable picture
whereby  young  early-type galaxies  form  in  a dissipative  collapse
process, and  then undergo increased  (either major or  minor) merging
activity in richer rather than in poor clusters.

\end{abstract}

\section{Introduction}
\label{INTR}
Early-type  galaxies (ETGs)  have negative  internal  color gradients,
with  their  stellar populations  becoming  bluer  towards the  galaxy
outskirts.   Different studies  have  shown that  these gradients  are
mainly caused  by a metallicity  variation (e.g.~\citealt{PVJ90}).  As
reviewed  by~\citet[hereafter PDI90]{PDI90},  color  gradients can  be
used as  an effective  discriminant among galaxy  formation scenarios.
In a monolithic  collapse picture, ETGs form by  the rapid collapse of
over-dense regions  at high redshift.  Galactic winds  are expected to
produce    steeper    metallicity    gradients   in    more    massive
galaxies~\citep{LAR74, CAR84}.  On the other hand, merging is expected
to produce  some dilution of the  gradients in galaxies~\citep{WHI80},
deleting  or  producing an  {\it  inverse}  correlation among  stellar
population  gradients and  luminosity, and  making the  gradients less
steep in denser environments.

It is  also unclear whether or  not color gradients in  ETGs depend on
environment and/or  galaxy luminosity.  \citet[hereafter TaO03]{TaO03}
found  the mean  color gradient  of  ETGs in  a nearby  cluster to  be
consistent  with  the  value  obtained  for field  galaxies  by  other
authors~(PDI90,~\citealt{IMP02}).   On  the other  hand,~\citet{TaO00}
and~\citet{TKA00} studied  color gradients  in both cluster  and field
galaxies at intermediate redshifts, and  found that the former seem to
have less steep  gradients.  To date, no study  of color gradients for
galaxies in groups has been performed.

~\citet{KDJ89}   examined  the  dependence   of  color   gradients  on
luminosity  and other galaxy  properties.  While  no clear  trend with
luminosity was found,  they noted that the largest  gradients occur in
intermediate  luminosity galaxies,  and suggested  that post-formation
mergers  on average  diminish  the primordial  gradients  in the  more
luminous  systems;  this was  also  supported  by  their finding  that
galaxies  with  more anisotropic  velocity  distributions have  weaker
color  gradients.  TaO03 also  showed that  for cluster  galaxies some
evidence exists for  a bimodal behavior, with very  bright ($L > L^*$)
ETGs  having  gradients  that  steepen with  luminosity,  and  fainter
galaxies showing  the opposite trend.~\citet{LBM04} and~\citet{dPCD04}
found that gradients in cluster ETGs  do not change or can become more
steep at fainter luminosities.

In this  Letter, we study the  dependence of $g-r$  color gradients in
ETGs  on  galaxy  luminosity  and environment.   Color  gradients  are
estimated  by  deriving structural  parameters,  namely the  effective
radius, $r_e$, and  the Sersic index, $n$, for a  sample of 1,700 ETGs
in 159  clusters\footnote{The list of  clusters is available  from the
first author.}  with different richnesses in the redshift range $ 0.05
\!  \lesssim  \!  z \!  \lesssim  \!  0.2$.  We  assume a $\Lambda$CDM
cosmology   with    $\Omega_m=0.3,\Omega_{\Lambda}=0.7,$   and   ${\rm
H_0}=70~{\rm km~s^{-1}~Mpc^{-1}}$.

\section{Data}
The data are taken from the Palomar Abell Cluster Survey~\citep{GAL00}
and  were collected  at  the 1.5m  telescope  at Palomar  with a  SITe
2048$\times$2048, AR-coated CCD.  The  pixel scale at this detector is
0.368 arcsec/pixel,  yielding a 12.56$^{\prime}\times$12.56$^{\prime}$
FOV.   Data were  taken in  ${\it g,r,i}$  $\rm Gunn  \!  -  \! Thuan$
filters.   We refer the  reader to~\citet{GAL00}  for more  details on
data reduction and photometric quality.

Galaxies used for this study were selected as follows~(see La Barbera
et al.~2005, in prep, for details).  We selected likely cluster
members as those galaxies whose $g-r$ colors are within $\pm \alpha
\cdot \sigma_{CM}$ around the color-magnitude relation of each
cluster,  where $\sigma_{CM}$ is  the observed  dispersion of  the red
sequence.   Field contamination  was estimated  using 34  blank fields
observed  with  the  same  configuration  and exposure  times  as  the
clusters. In  order to keep  field contamination as small  as possible
without overly reducing the total number of ETGs, we use $\alpha \!  =
\!  0.5$.   From the 34  blank fields, we  measure the mean  number of
field galaxies  found in  the same magnitude  and color range  used to
select galaxies  in each of the clusters.   Choosing $\alpha=0.5$, the
fraction  of field  galaxies was  typically $\sim  \! 10  \%$  for the
entire cluster sample.  We note  that decreasing the value of $\alpha$
reduces  dramatically  the  total  number  of  ETGs  without  changing
significantly   the  field   contamination.   For   example,  choosing
$\alpha=0.25$ decreases the number of ETGs by $40
\%$, while the  fraction of expected field galaxies  decreases by only
$1 \%$.   On the  other hand, increasing  the value of  $\alpha$ makes
field  contamination  significantly  higher.   For  example,  choosing
$\alpha=0.75, 1.0$ or $1.5$,  the fraction of field galaxies increases
to $16$, $18$ or $20\%$  respectively.  

The value of $\sigma_{CM}$ was obtained for each cluster by fixing the
red  sequence   slope  to  the  value  given   by~\cite{VS77}  at  the
corresponding redshift and by  computing a $3 \sigma$ clipped standard
deviation of residuals to the red sequence.  We excluded galaxies with
overlapping isophotes, which were  defined using an isophotal level of
1.5$\sigma$ of the background  standard deviation. These objects could
be  interacting  galaxies  at  the  cluster  redshift  or  objects  at
different redshifts that overlap  because of projection effects.  This
selection minimizes cases in which  color gradients might be driven by
interaction with  nearby objects rather than by  global cluster and/or
galaxy properties.   For each cluster, we also  selected only galaxies
brighter than a given magnitude  limit.  We simulated galaxy images as
a function of  redshift and calculated the necessary  $S/N$ to recover
the $g-r$ color  gradients with systematic errors less  than 0.005 mag
(~\citealt[hereafter  LBM02]{LBM02}).  For  each cluster,  we  set the
magnitude limit such  that this $S/N$ is achieved  in the radial range
used  (see  below).  The  typical  value  for  the magnitude  cut  was
$r=18.5$.

The  above  criteria  yield  a  sample of  1,950  galaxies  for  which
structural parameters  were derived  in the $g$  and $r$  bands.  From
these galaxies, we selected the  1,700 objects with Sersic index $n \!
> \!   2.2$, which  are likely  cluster ETGs~\citep{BHB03}.   For each
cluster, we measure the richness from DPOSS galaxy catalogs
\citep{gal03}. Only 108 of the clusters (containing 1,135 ETGs) have
well-calibrated DPOSS data. Additionally, a number of systematic
effects could introduce a bias into our results at higher
redshift. These effects include the enhanced detectability of rich
clusters as opposed to poor ones, the faintness and smaller sizes of
the galaxies in high-$z$ clusters, and sampling of a different portion
of the rest-frame galaxy spectra in the $g$ and $r$
filters. Therefore, for the richness dependent analyses we further
restrict the dataset to the 560 galaxies in 75 well-measured clusters
with $z<0.15$.

\section{Derivation of Color gradients }
\label{MET}
Surface  photometry was  derived  by fitting  galaxy  stamps with  PSF
convolved  Sersic models,  as detailed  in LBM02.   For each  stamp, a
constant background  value was  fitted simultaneously with  the model,
achieving a typical accuracy of better than $0.1 \%$ in the background
value.   For  each cluster  and  in each  filter,  the  PSF model  was
constructed  by  fitting  star  images  with a  sum  of  three  Moffat
functions,  giving a  mean  value of  $\sim  \! 1  $  for the  reduced
$\chi^2$ of the  PSF fits in the whole sample.   Deviations of the PSF
from  a circular  shape  were  also taken  into  account by  expanding
stellar isophotes in a cos and  sin series. Details will be given in a
forthcoming  paper~(La  Barbera  et  al.~2005, in  preparation).   The
best-fitting structural parameters ($\rm r_e$  and $n$) in the $g$ and
$r$  bands were used  to derive  the internal  color profile  for each
galaxy, $g-r(\rm  r)$, where $\rm r$  is the distance  from the galaxy
center.  The color gradient, defined as $\nabla_{g-r} = \delta (g-r) /
\delta(\log \rm  r)$, was  estimated by the  logarithmic slope  of the
color profile, performing a linear  fit of $g-r(\rm r)$ vs.  $\log \rm
r$.   For comparison  with  other works  (e.g.~PDI90),  the fits  were
performed in the range $\rm r_{min}  < \rm r < \rm r_{max}$, with $\rm
r_{min} =  0.1 r_e$ and  $\rm r_{max} =  r_e$, where $\rm r_e$  is the
effective radius in the $r$ band.   The typical value of $\rm r_e$ for
the galaxies  analyzed here varies from  $\sim \! 1.9$\arcsec  at $z <
0.15$ to  $\sim 1.1$\arcsec  at higher redshifts.   We also  note that
slightly varying (by $\sim 30  \%$) the values of $\rm r_{min}$ and/or
$  \rm  r_{max}$  changes  the~\cg  values by  less  than  $0.01$~mag,
demonstrating the robustness of our results.

\begin{figure}
\epsscale{0.75}
\plotone{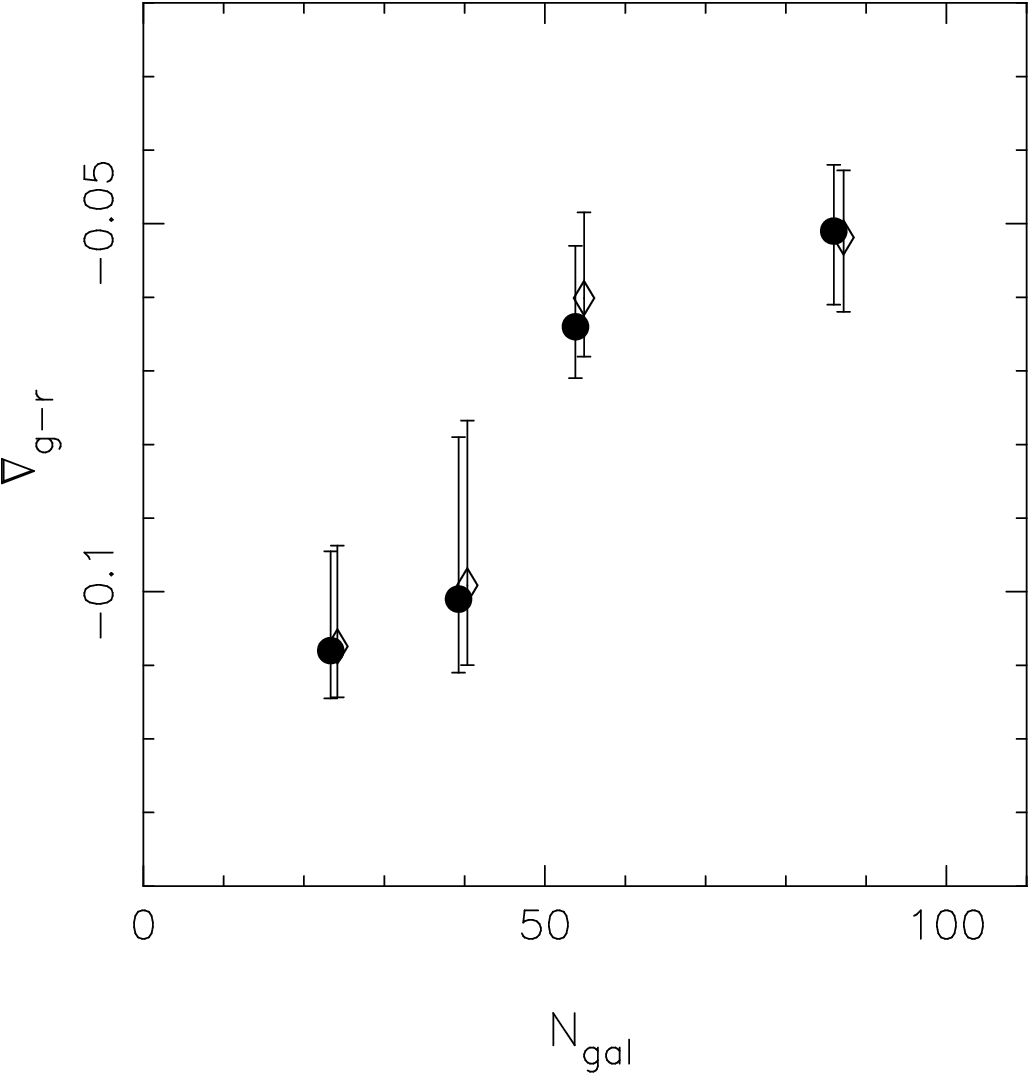}
\caption{For clusters with $z<0.15$ and well measured richnesses, we
    plot the mean value of  the $g-r$ color gradients, \cg~ versus the
mean cluster richness, $N_{gal}$, (filled symbols).  Each bin includes
$N=140$ galaxies. Empty circles are the mean values of
\cg \, corrected  for field contamination.   Error bars denote  $1 \sigma$
uncertainties.  A small offset in  richness has been added to separate
the corrected and uncorrected data points. }
\label{figR}
\end{figure}
\section{Results}
\label{RES}
Galaxies  were binned  according to  their parent  cluster richnesses,
each bin having the same number  of galaxies.  As noted in \S2, we use
only the $N=560$ galaxies in clusters  at $z \!  < \!  0.15$, which is
the median  redshift of the  whole sample.  Fig.~\ref{figR}  shows the
mean color gradient (filled symbols) as a function of the mean cluster
richness N$_{\rm gal}$.  The  figure clearly shows that~\cg \, becomes
less steep for galaxies in richer clusters.  The difference in ~\cg \,
between points  with N$_{\rm gal}  < 50$ and  N$_{\rm gal} > 50$  is $
-0.047 \pm  0.008$, and is  significant at $~5.9\sigma$.  In  order to
estimate the effect of field  contamination on the~\cg \, vs.  N$_{\rm
gal}$ relation, we used a Monte-Carlo technique.  For each cluster, we
measure $N_c$  galaxies which are likely  to be ETGs.   From the blank
fields we  then measure  $N_f$ galaxies which  follow the same  set of
criteria for  being in  the cluster -  these are the  contaminants. To
account  for possible  magnitude-dependent differences  between galaxy
populations,  the cluster and  field galaxies  are divided  into three
magnitude bins,  and in  each bin  $i$ we randomly  select a  total of
$N_{c,i} - N_{f,i}$ distinct galaxies which are expected to be cluster
members.   This  procedure was  iterated  500  times  for all  of  the
clusters, re-computing  at each iteration  the mean color  gradient in
each  richness range.   Fig.~\ref{figR} plots  the mean  value  of the
corrected~\cg \, values as  the open symbols.  The corresponding error
bars are  obtained by adding  in quadrature the standard  deviation of
the  background-corrected   ~\cg\,  values   to  the  errors   on  the
uncorrected  gradients.  The difference  between the  corrected points
with N$_{\rm gal} < 50$ and N$_{\rm gal} > 50$ is $ -0.047 \pm 0.009$,
which  is significant  at $\sim  5.5\sigma$.  We  conclude  that field
contamination has no effect on our results.
\begin{deluxetable}{cccc}
\tablecaption{Mean color gradients for galaxies in  clusters with different richnesses}
\tablewidth{0pt}
\tablecolumns{4}
\tablehead{
\multicolumn{2}{c}{LOW R} & \multicolumn{2}{c}{HIGH R} \\
\colhead{z} & \colhead{$\nabla_{g-r}$}  & \colhead{z} & \colhead{$\nabla_{g-r}$} }
\startdata
  0.08 &  $-0.064_{0.02}^{0.008}$ & 0.105  & $-0.04_{0.013}^{0.012}$\\
  0.12 & $-0.106_{0.02}^{0.015}$  & 0.14 & $-0.056_{0.013}^{0.012}$ \\
  0.16 &  $-0.103_{0.013}^{0.02}$ & 0.19  & $-0.08_{0.026}^{0.024}$ \\
  0.21 & $-0.137_{0.014}^{0.026}$ & 0.23 & $-0.079_{0.027}^{0.026}$ \\
\enddata
\label{cg_z_R}
\end{deluxetable}

\begin{figure}
\epsscale{0.75}
\plotone{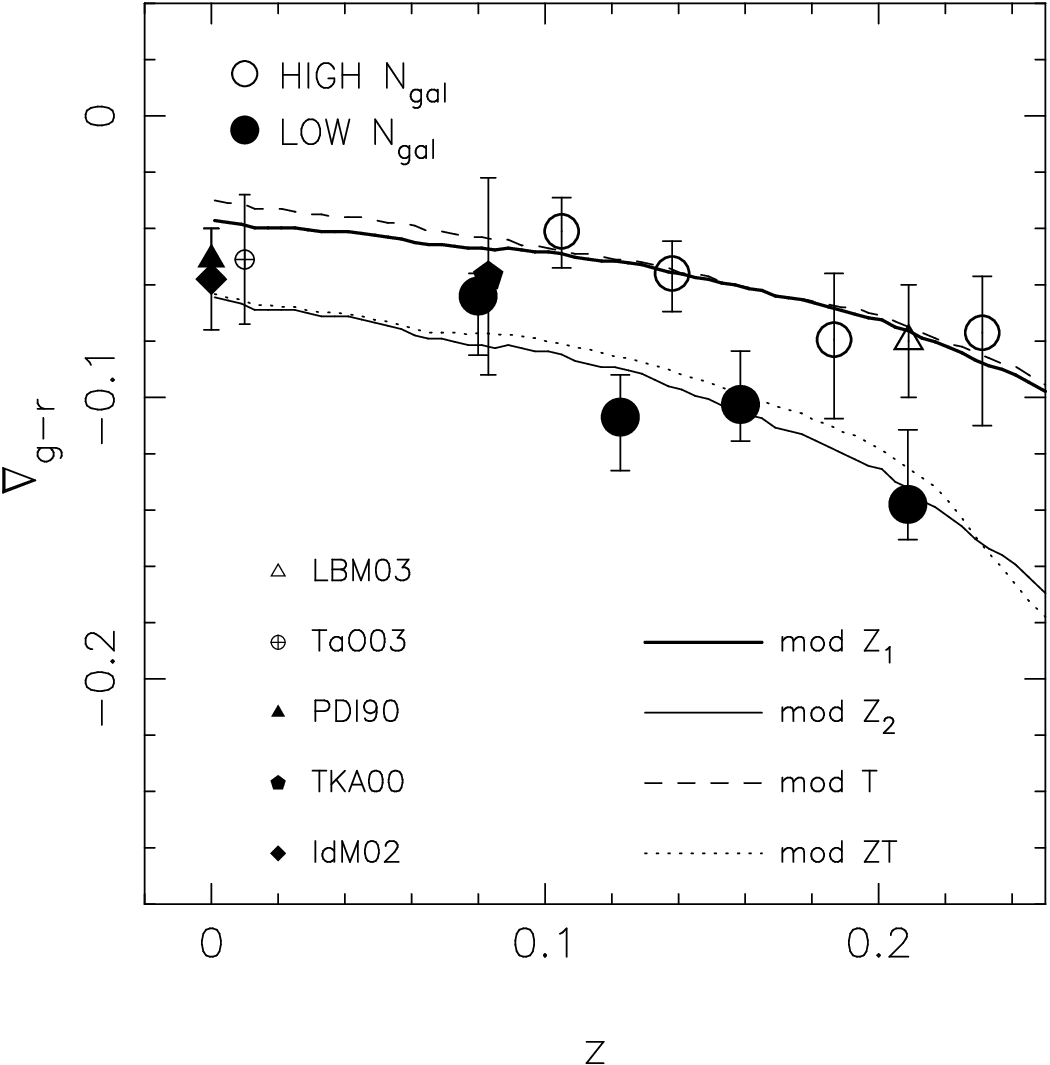}
\caption{The mean  $g-r$ color gradient, \cg, is plotted versus redshift for
high   and  low   richness   clusters  (empty   and  filled   circles,
respectively).  Error bars denote $1 \sigma$ uncertainties.  Gradients
from the literature  for field and cluster ETGs  are plotted as filled
and  empty  symbols  respectively,  (see references  in  the  figure).
Curves  represent   the  stellar  population  models   for  the  color
gradients.}
\label{figZ}
\end{figure}
Fig.~\ref{figZ} shows the  mean value of \cg \,  in different redshift
bins.  The whole  sample was divided into low  ($N_{gal}<50$) and high
($N_{gal}>50$)  richness  clusters  (hereafter  LRCs and  HRCs),  with
similar  numbers of galaxies  in both  subsamples.  The  gradients for
each subsample are given in Table~\ref{cg_z_R}.  In Fig.~\ref{figZ} we
also show optical color gradients from previous works for field (empty
symbols) and cluster  (filled symbols) ETGs at $z  \lesssim 0.2$.  The
gradients  were  transformed  to~\cg  \,  values  using  the  GISSEL03
spectral code~\citep{BrC03}, as detailed in La Barbera et al.~2005 (in
preparation).   We  see that  our  data  are  in good  agreement  with
previous measurements of  color gradients.  Fig.~\ref{figZ} shows that
color  gradients  slightly   decrease  with  redshift,  following  two
distinct trends depending on  the cluster richness.  To quantify these
trends, we  performed a linear fit of  ~\cg \, vs.  $z$  for both LRCs
and HRCs, using a least squares procedure with~\cg \, as the dependent
variable. The uncertainties on  the fitting coefficients were obtained
by shifting  points in Fig.~\ref{figZ} according  to the corresponding
error bars  and repeating  the fitting procedure.   The slopes  of the
fitted lines are $-0.52 \pm 0.18$ for LRCs and $-0.32
\pm 0.2$ for HRCs.  These values are negative at $2.9$ and $1.6\sigma$
significance levels, respectively.  We note that the difference
between~\cg \, of LRCs and HRCs in Tab.~\ref{cg_z_R} is $-0.043 \pm
0.008$, which is significant at $\sim 5\sigma$ (cfr.  above). 

To  constrain the  underlying  age and  metallicity  gradients in  the
galaxies, we  used four empirical  stellar population models  for \cg:
two pure metallicity models ($Z_1$ and $Z_2$), a pure age model ($T$),
and a mixed age+metallicity  model ($TZ$).  Each model was constructed
by  computing  the difference  in  color  between  two simple  stellar
populations, describing  the properties of stellar  populations at the
inner and  outer galaxy radii,  $\rm r_{min}$ and $\rm  r_{max}$, used
for the  computation.  Galaxy colors  were obtained from  the GISSEL03
code.  The  inner population was  assumed to be old  ($T=12$~Gyr) with
solar  metallicity,  while  the  age  and  metallicity  of  the  outer
population were  changed to reproduce the actual  color gradients.  In
models  $Z_1$ and  $Z_2$  the  outer populations  are  old, but  their
metallicities are changed in order  to describe the color gradients in
HRCs and LRCs,  respectively.  Model $TZ$ was created  to describe the
gradients  in  poorer  clusters  by  adding an  age  gradient  to  the
metallicity  gradient  of  model  $Z_1$, which  reproduces  the  color
gradient  in richer  clusters.   In  model $T$,  the  inner and  outer
stellar populations have the  same metallicity, while the age gradient
is  chosen   to  reproduce  the  color  gradient   in  rich  clusters.
Table~\ref{MOD}  reports  the   relevant  parameters  of  each  model,
together  with  the   corresponding  metallicity  and  age  gradients,
$\nabla_Z=   \log   (Z_o/Z_i)$  and   $\nabla_T   =  \log   (T_o/T_i)$
respectively.
\begin{deluxetable}{ccccccc}
\tabletypesize{\scriptsize}
\tablecolumns{7}
\tablewidth{0pc}
\tablecaption{Parameters of stellar population models for color gradients 
}
\tablehead{
\colhead{model} & $Z_i/Z_\odot$ & $T_i$(Gyr) & $Z_o/Z_\odot$ &
$T_o$(Gyr) & $\nabla_Z$ & $\nabla_T$ }
\startdata
 $Z_1$ & $1.0$ & $11.4$ & $0.6 $ & $11.4$ & $-0.22$  & $0$ \\
 $Z_2$ & $1.0$ & $11.4$ & $0.4 $ & $11.4$ & $-0.4$ & $0$ \\
 $TZ$  & $1.0$ & $11.4$ & $0.6 $ & $8.0$ & $-0.22$ & $-0.15$ \\
 $T$   & $1.0$ & $11.4$ & $1.0  $ & $8.0$ & $0$ & $-0.15$ \\
\enddata
\label{MOD}
\end{deluxetable}

\begin{figure}
\epsscale{0.75}
\plotone{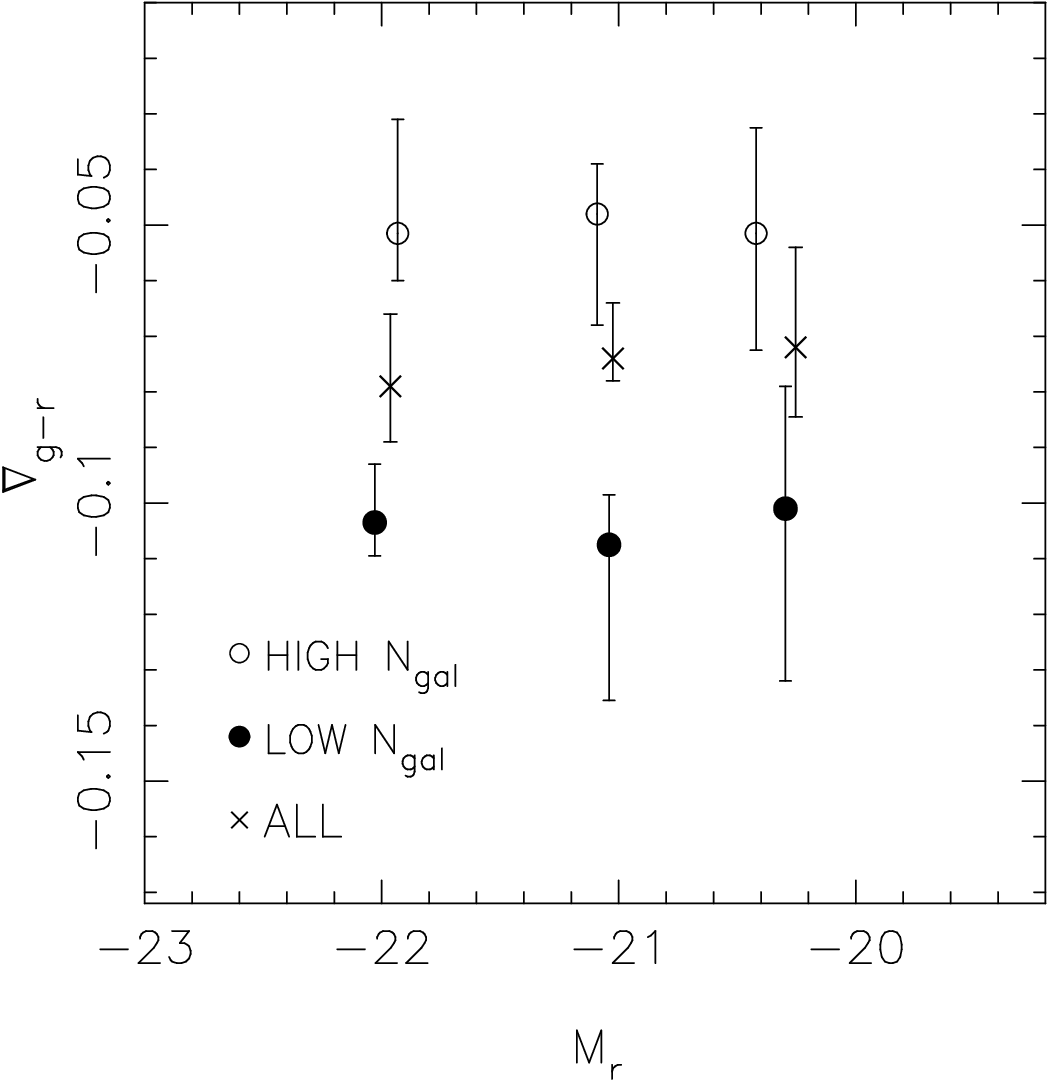}
\caption{Mean values of \cg \, versus galaxy absolute magnitude, $M_r$. 
Different symbols are used for HRCs, LRCs and for the whole sample, as
shown in the lower left part of the figure.  Error bars denote $1
\sigma$ uncertainties.}
\label{figM}
\end{figure}
Fig.~\ref{figM} plots  the mean value of  \cg \, for galaxies  at $z <
0.15$  as  a function  of  their  absolute  $r$ magnitude,  which  was
obtained  by  using k-  corrections  from  the  GISSEL03 code.   Three
samples are shown  in the plot, corresponding to  galaxies in clusters
with low, high,  and all richnesses,  respectively. We see that
color gradients do not depend significantly on galaxy luminosity.

\section{Discussion}
The dependence of color gradients on cluster richness can be explained
by  the   fact  that  some  physical  mechanism,   related  to  galaxy
environment,  affects the age  and metallicity  gradients in  ETGs.  A
pure  metallicity   gradient  can  account  for   the  observed  color
gradients,  provided that~\cz \,  flattens from  about $  -0.4$ (model
$Z_2$)  in  LRCs  to  $\sim   -0.2$  (model  $Z_1$)  in  HRCs.   These
metallicity gradients can be compared with those estimated by previous
works,  which found  them  to be  in  the range  $  \left[ -0.3,  -0.2
\right]$,    with     a    typical    uncertainty     of    $0.1$~(see
e.g.~\citealt{IMP02}).  Both models $Z_1$ and $Z_2$, therefore, are in
good agreement with previous~\cz \, estimates.  The presence of overly
shallow  metallicity gradients  in ETGs  has often  been invoked  as a
serious problem  for monolithic  scenarios of galaxy  formation, since
these models predict~\cz \, values in the range $
\left[ -1, -0.3 \right] $ (\citealt{LAR74}, \citealt{CAR84}, \citealt{KAW01},
hereafter KAW01).   Models $Z_1$ and  $Z_2$ show, however,  that color
gradients in  galaxies in LRCs are fully  consistent with expectations
from the  monolithic formation scenario,  but the consistency  is only
marginal for HRCs.  A natural  candidate for the variation of the mean
color gradient as  a function of environment is  galaxy merging, which
is   expected  to   flatten   the  stellar   population  gradient   in
galaxies~\citep{WHI80}. Our  results support a scenario  in which ETGs
have undergone  increased (either major  or minor) merger  activity in
richer rather  than in  poor clusters. Models  $T$ and  $TZ$, however,
show that this interpretation is not unique.  For example, the results
for  rich clusters  can be  equally  well described  by a  metallicity
gradient of $-0.22$ or by an  age gradient of $-0.15$ (model $T$). The
ratio of  these gradients  is $\nabla_Z /  \nabla_T \! \sim  \!  3/2$,
which     is     simply     the     well     known     age-metallicity
degeneracy~\citep{WTF96}.    Model  $TZ$   shows  that   the  richness
dependence can also be explained by fixing the metallicity gradient of
both HRCs and  LRCs to that of model $Z_1$,  provided that galaxies in
LRCs have a  younger stellar population in the  outskirts, with an age
gradient of about $-0.15$.  We note that a mild age gradient, of about
$-0.1$, would  also be consistent  with previous studies~\citep{SMG00,
LBM03}.  In a hierarchical merging picture, galaxies tend to develop a
disk by accreting  gas from the surrounding halo,  with this mechanism
being inhibited in  high density environments~\citep{KWG93}.  It would
be conceivable,  therefore, that a negative age  gradient accounts for
the steepening of the observed color gradients in LRCs.

A troublesome  issue for galaxy  formation theories is the  absence of
any sharp  correlation between  color gradient and  galaxy luminosity.
In a  monolithic formation picture,  the same galactic  wind mechanism
which  is required  to explain  the color-magnitude  relation  of ETGs
would  produce  a  steepening  of  the metallicity  gradient  in  more
luminous galaxies (KAW01).  Following KAW01 (model B in their Tab.~3),
we expect a \cg \, change  of about $-0.08$ in the luminosity range of
Fig.~\ref{figM},  which is in  clear disagreement  with our  data.  We
note that a steepening of the  color gradients as a function of galaxy
luminosity is expected if ETGs form through the dissipative merging of
disk  systems~\citep{BeS99}.   The   absence  of  correlation  between
gradient   and  luminosity   could  be   explained  by   invoking  (i)
dissipationless merging,  which could be more  effective in flattening
the metallicity gradient of massive  galaxies, or (ii) a steepening of
age  gradients in less  massive galaxies,  which could  counteract the
metallicity-luminosity  relation.  It is  not obvious,  however, which
physical  mechanism  could  drive  this  antagonistic  behavior.   The
present  data show  that  internal color  gradients  in ETGs  probably
result from  a complex  coordination of different  physical processes.
Crucial insights  into these processes  could be obtained  by breaking
the  age-metallicity degeneracy, exploring  color gradients  for large
samples  of  galaxies  in  the  optical--NIR wavebands  or  at  higher
redshift.

\acknowledgements
RdC,  RRG,  and  SGD  acknowledge  partial  support  from  the  Norris
Foundation  and the  Ajax  Foundation.  We  also  thank the  anonymous
referee whose comments have helped us to improve the paper.  This work
has  been partially supported  by the  Italian Ministry  of Education,
University, and Research (MIUR) grant COFIN2004020323.

\end{document}